\documentclass[twocolumn,showpacs,preprintnumbers,amsmath,amssymb,aps]{revtex4}
\usepackage{graphicx}% Include figure files
\usepackage{dcolumn}% Align table columns on decimal point
\usepackage{bm}% bold math

\def\ket#1{\mathinner{|{#1}\rangle}}

\begin{document}

\title{Role of population transfer under strong probe conditions in electromagnetically induced transparency}
 \author{Kanhaiya Pandey}
 \author{Dipankar Kaundilya}
 \author{Vasant Natarajan}
 \email{vasant@physics.iisc.ernet.in}
 \homepage{www.physics.iisc.ernet.in/~vasant}
 \affiliation{Department of Physics, Indian Institute of
 Science, Bangalore 560\,012, INDIA}

\begin{abstract}
We analyze theoretically the phenomenon of
electromagnetically induced transparency (EIT) under
conditions where the probe laser is not in the usual weak
limit. We consider the effects in both three-level and
four-level systems, which are either closed or open (due to
losses to an external metastable level). We find that the
EIT dip almost disappears in a closed three-level system
but survives in an open system. In four-level systems,
there is a narrow enhanced-absorption peak (EITA) at line
center, which has applications as an optical clock. The
peak converts to an EIT dip in a closed system, but again
survives in an open system.
\end{abstract}

%42.50.Gy Effects of atomic coherence on propagation, absorption, and amplification of light;
%         electromagnetically induced transparency and absorption
%42.50.Nn Quantum optical phenomena in absorbing, amplifying, dispersive and conducting media;
%         cooperative phenomena in quantum optical systems
%32.80.Qk Coherent control of atomic interactions with photons

\maketitle

\section{Introduction}
The phenomenon of electromagnetically induced transparency
(EIT) is generally studied in multilevel systems under
conditions where a {\it strong} control laser drives one
transition and thereby modifies the absorption properties
of the medium for a {\it weak} probe laser on another
transition. The modification happens because the control
laser shifts the energy levels (due to the AC-Stark effect
resulting in the formation of ``dressed'' states) and the
subsequent quantum interference among the new pathways for
probe absorption. In addition, the control laser, when it
couples to a ground level, causes population redistribution
among the levels due to optical pumping. The probe laser is
considered weak enough that it does not play a role in
either shifting the energy levels or causing population
transfer. EIT has important applications in diverse fields
such as lasing without inversion \cite{IMH89,AGA91},
high-resolution spectroscopy \cite{RAN02,KPW05},
enhancement of second-order and third-order nonlinear
processes \cite{HFI90,ZKA09}, polarization control
\cite{WIG98,PWN08} and slowing of light \cite{HHD99}. In
many of these applications, particularly when studying
nonlinear processes, the weak-probe assumption is not
valid. There are also recent proposals
\cite{HCN05,SAI05,TYO06} to use control lasers in a
four-level system to drive an optical clock transition. The
probe laser is on a weakly-allowed intercombination line.
Our experiments using a similar transition in Yb have shown
that the probe laser must be operated near saturation
intensity to get good signal-to-noise ratio from the weak
transition.

We thus see that it is important to understand the role of
a {\it strong} probe laser in EIT experiments. There have
been two previous studies using a strong probe laser in a
ladder-type system in Rb \cite{WIG98a,PAN08}. This system
can be considered closed since there are no decay pathways
out of the three levels under consideration. However, in
atoms such as Sr and Yb, the presence of low-lying
metastable states makes them open or lossy. We therefore
study the modification to probe absorption due to a strong
probe laser in both closed and open systems. As we will
see, the main effect of the strong probe is in causing
additional population transfer between the levels. We also
consider four-level systems due to their importance in
optical-clock proposals.

\section{Density-matrix analysis}
\begin{figure}
\centering{\resizebox{0.9\columnwidth}{!}{\includegraphics{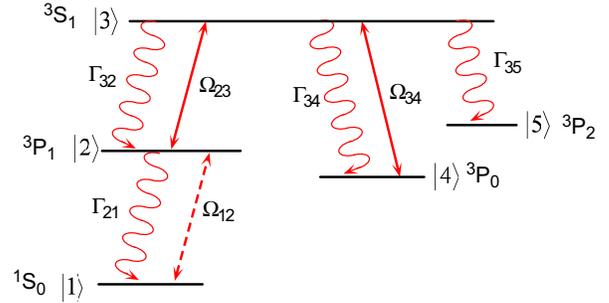}}}
\caption{Multilevel system under consideration. Shown
alongside each level is the atomic state that can be used
to realize this level scheme in Yb.}
 \label{4level}
\end{figure}

Consider the multi-level system shown in Fig.\
\ref{4level}. Levels $\ket{4}$ and $\ket{5}$ are metastable
levels, so that atoms decaying to these levels are lost
unless they are repumped. The probe laser is on the
$\ket{1} \leftrightarrow \ket{2}$ transition, while there
are two control lasers, on the $\ket{2} \leftrightarrow
\ket{3}$ transition and the $\ket{3} \leftrightarrow
\ket{4}$ transition, respectively. The Rabi frequency of a
laser driving the $\ket{i} \leftrightarrow \ket{j}$
transition is $\Omega_{ij}$ and its detuning from resonance
is $\Delta_{ij}$. The corresponding decay rate from the
upper level $\ket{j}$ to the lower level $\ket{i}$ is
$\Gamma_{ji}$. This level scheme can be made three- or
four-level, and open or closed, by choosing suitable
parameters to be 0. For example, if $\Omega_{34}$ is 0, it
becomes a three-level ladder system. Simultaneously, if
$\Gamma_{34}$ and $\Gamma_{35}$ are 0, the system becomes
closed, and open otherwise. The general four-level system
is again closed if $\Gamma_{35}$ is 0, and open otherwise.

The time evolution of the system in the rotating-wave
approximation is given by the following set of
density-matrix equations. As usual, $\rho_{ii}$ gives the
population in level $\ket{i}$, while $\rho_{ij}$ gives the
coherence between levels $\ket{i}$ and $\ket{j}$.
\begin{eqnarray}
 \rho_{11} &=& \Gamma_{21}\rho_{22}+\frac{i}{2} \left[ \Omega_{12}\rho_{21}-\Omega^*_{12}\rho_{12} \right]  ,\nonumber \\
 \rho_{22} &=& -\Gamma_{21}\rho_{22}+\Gamma_{32}\rho_{33}+\frac{i}{2}\left[ \Omega^*_{12}\rho_{12}-\Omega_{12}\rho_{21} \right] \nonumber \\
 & & +\frac{i}{2}\left[ \Omega^*_{23}\rho_{32}-\Omega_{23}\rho_{23} \right]  ,\nonumber \\
 \rho_{33} &=& -\Gamma_{32}\rho_{33}-\Gamma_{34}\rho_{33}-\Gamma_{35}\rho_{33}\nonumber \\
 & & +\frac{i}{2}\left[ \Omega^*_{23}\rho_{23}-\Omega_{23}\rho_{32}\right] +\frac{i}{2}\left[ \Omega_{34}\rho_{43}-\Omega^*_{34}\rho_{34}\right]  ,\nonumber \\
 \rho_{44} &=& \Gamma_{34}\rho_{33}-\Gamma_{41}\rho_{44}+\frac{i}{2}\left[ \Omega^*_{34}\rho_{34}-\Omega_{34}\rho_{43}\right]  ,\nonumber \\
 \rho_{12} &=& -\left[ \frac{\Gamma_{21}}{2}+i\Delta_{12}\right] \rho_{12}+\frac{i}{2}\left[ \Omega_{12}(\rho_{22}-\rho_{11})\right] \nonumber \\
 & & -\frac{i}{2}\left[ \Omega^*_{23}\rho_{13}\right] ,\nonumber \\
 \rho_{13} &=& -\left[ \frac{\Gamma_{32}}{2}+\frac{\Gamma_{34}}{2}+\frac{\Gamma_{35}}{2}+i(\Delta_{12}+\Delta_{23})\right] \rho_{13}\nonumber \\
 & & +\frac{i}{2}\left[ \Omega_{12}\rho_{23}-\Omega_{23}\rho_{12}-\Omega^*_{34}\rho_{14}\right]  ,\nonumber \\
 \rho_{14} &=& -\left[ \frac{\Gamma_{41}}{2}+i(\Delta_{12}+\Delta_{23}-\Delta_{34})\right] \rho_{14}\nonumber \\
 & & +\frac{i}{2}\left[ \Omega_{12}\rho_{24}-\Omega_{34}\rho_{13}\right]  ,\nonumber \\
 \rho_{23} &=& -\left[ \frac{\Gamma_{32}}{2}+\frac{\Gamma_{34}}{2}+\frac{\Gamma_{35}}{2}+\frac{\Gamma_{21}}{2}+i\Delta_{23}\right] \rho_{23}\nonumber \\
 & & +\frac{i}{2}\left[ \Omega_{12}\rho_{13}+\Omega_{23}(\rho_{33}-\rho_{22})-\Omega^*_{34}\rho_{24}\right]  ,\nonumber \\
 \rho_{24} &=& -\left[ \frac{\Gamma_{21}}{2}+\frac{\Gamma_{41}}{2}+i(\Delta_{23}-\Delta_{34})\right] \rho_{24}\nonumber \\
 & & +\frac{i}{2}\left[ \Omega_{12}\rho_{14}+\Omega_{23}\rho_{34}-\Omega_{34}\rho_{23}\right] ,\nonumber \\
 \rho_{34} &=& -\left[ \frac{\Gamma_{32}}{2}+\frac{\Gamma_{34}}{2}+\frac{\Gamma_{35}}{2}+\frac{\Gamma_{41}}{2}-i\Delta_{34}\right] \rho_{34}\nonumber \\
 & & +\frac{i}{2}\left[ \Omega^*_{23}\rho_{24}+\Omega_{34}(\rho_{44}-\rho_{33})\right] .
 \label{density}
\end{eqnarray}

The usual method of solving these equations in the weak
probe limit is to look for a steady-state solution {\it to
first order in the probe Rabi frequency}. Since we want to
see the modification due to a strong probe laser, we solve
these equations numerically with arbitrary values of the
Rabi frequency. The other parameters are taken from the
relevant level scheme in Yb, i.e., $\Gamma_{21} = 2\pi
\times 0.18$ MHz, $\Gamma_{32} = 2\pi \times 5.89$ MHz,
$\Gamma_{34} = 2\pi \times 1.54$ MHz, and $\Gamma_{35} =
2\pi \times 4.29$ MHz \cite{LBM02}. The metastable state
$\ket{4}$ in Yb has a lifetime of order months, hence the
decay rate $\Gamma_{41}$ is taken to be 0.

\begin{figure}
\centering{\resizebox{0.95\columnwidth}{!}{\includegraphics{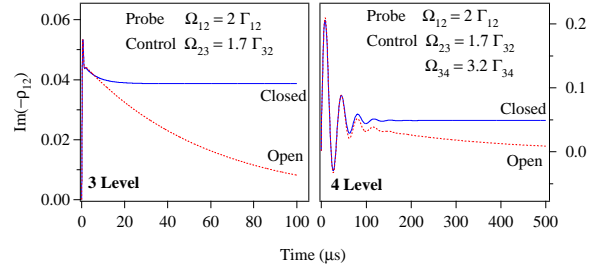}}}
\caption{Time evolution of Im($-\rho_{12}$) for open and
closed systems. For open systems, there is no steady state
and the amplitude continues to decay.}
 \label{time}
\end{figure}

Of course, for open systems, there is no true steady state,
since all the population will be lost to the metastable
levels as $t \rightarrow \infty$. In Fig.\ \ref{time}, we
show the time evolution of Im($-\rho_{12}$) under typical
strong probe conditions for both open and closed systems.
In the three-level case, the transients for a closed system
last a few microseconds and then stabilize to the steady
state value. In an open system, the transients last for the
same time, after which the amplitude continues to decay but
without any no structure. Hence our results are given after
100~$\mu$s. For four-level systems, the transients in both
systems last for almost 200~$\mu$s. Therefore, the results
are given after 500~$\mu$s.

\section{Results in a three-level ladder system}
We first consider the effect of a strong probe in a
three-level ladder system. As explained earlier, we realize
this system by setting $\Omega_{34}=0$. We compare the
effect of the probe power in both a closed system (by
setting $\Gamma_{34} = \Gamma_{35} =0$) and an open system
(with just $\Omega_{34}=0$).

In the density-matrix analysis, probe absorption is
proportional to Im($-\rho_{12}$), where $\rho_{12}$ is the
induced absorption on the $\ket{1} \leftrightarrow \ket{2}$
transition coupled by the probe laser. In the weak-probe
limit, we can assume that all the population is in the
ground level, i.e., $\rho_{11} = 1$, and $\rho_{22} =
\rho_{33} = 0$. This assumption breaks down when the probe
laser is strong. If we do a series expansion in probe Rabi
frequency, the solution to the density-matrix equations
yields,
\begin{equation}
\rho_{12}=-\frac{\frac{i}{2}\frac{\Omega_{12}}{\gamma_{12}}\left[ 1+A \right]
- \frac{i}{8}\frac{\Omega_{23}^2\Omega_{12}}{\gamma_{12}\gamma_{13}\gamma_{23}}\left[ A \right] \left[ 1+B \right] }
{1+\frac{\Omega_{23}^2}{4\gamma_{12}\gamma_{13}}\left[ 1+B \right] },
\end{equation}
where
\begin{eqnarray}
\gamma_{12}&=&\Gamma_{21}/2+i\Delta_{12} \, , \nonumber \\
\gamma_{13}&=&(\Gamma_{32} + \Gamma_{34} +\Gamma_{35})/2 +i(\Delta_{12}+\Delta_{23}) \, , \nonumber \\
\gamma_{23}&=&(\Gamma_{21} + \Gamma_{32} + \Gamma_{34} +\Gamma_{35})/2 +i\Delta_{23} \, , \nonumber
\end{eqnarray}
and the two variables
\begin{eqnarray}
A &=& \left[ -\frac{1}{2}\frac{\Omega_{12}^2}{\Gamma_2\gamma_{12}}+\ldots\right] , \nonumber \\
{\rm and \ \ }
B &=& \left[ -\frac{1}{4}\frac{\Omega_{12}^2}{\gamma_{13}\gamma_{23}}
+\frac{1}{16}\frac{\Omega_{12}^4}{\gamma_{13}^2\gamma_{23}^2}+ \ldots \right] \, , \nonumber
\end{eqnarray}
arise from the modified population differences. As
expected, the expression for $\rho_{12}$ reduces to the
weak-probe expression when $A=B=0$.

\begin{figure}
\centering{\resizebox{0.95\columnwidth}{!}{\includegraphics{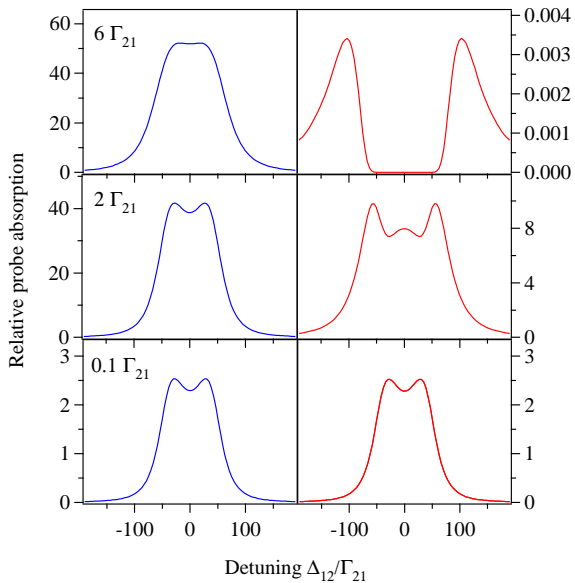}}}
\caption{Relative probe absorption [defined as
Im($-\rho_{12}$)] plotted as a function of probe detuning
for three values of probe Rabi frequency $\Omega_{12}$.
Both parameters are measured in units of $\Gamma_{21}$. The
strength of the control laser is $\Omega_{23}=2\pi \times
10$~MHz. The left-hand side is for a closed system and the
right-hand side is for an open system.}
 \label{threelevel}
\end{figure}

The numerical solutions of Im($-\rho_{12}$) for three-level
closed and open systems at three values of probe Rabi
frequency ($\Omega_{12}$) are shown in Fig.\
\ref{threelevel}. The control laser is taken to have a
strength of $\Omega_{23} = 2\pi \times 10$~MHz ($1.7 \,
\Gamma_{32}$). To make the comparison reasonable for the
two cases, the decay rate from $\ket{3} \rightarrow
\ket{2}$ for the closed system is taken to be
$\Gamma_{32}+\Gamma_{34}+\Gamma_{35}$, and the decay rate
from $\ket{3} \rightarrow \ket{4}$ for the open system is
taken to be $\Gamma_{34}+\Gamma_{35}$. This ensures that
the decay term governing $\rho_{13}$ is the same in both
cases. Both solutions are given at $t=100$~$\mu$s.

Let us first consider the results for the closed system
shown on the left-hand side. When $\Omega_{12}$ is very
small (the weak-probe limit of $0.1 \, \Gamma_{21}$), we
see the usual EIT dip at line center. When the power in
increased to $2\, \Gamma_{21}$, the EIT dip decreases,
which can be interpreted as arising due to interference
between new absorption paths created by the probe laser
that were negligible when the laser was weak. This decrease
continues with increased probe power so that, at $6\,
\Gamma_{21}$, the EIT dip almost completely disappears. For
open systems, shown on the right-hand side of the figure,
population loss into metastable levels is not significant
in the weak-probe limit, as expected. When the probe power
is increased, population loss causes the overall absorption
amplitude to go down. However, competition between optical
pumping and EIT allows the EIT dip to survive but with a
split line shape. Such a splitting of the EIT line has also
been observed before by us under strong-probe conditions in
a {\it closed} ladder system, but in that case it arose due
to thermal averaging \cite{PAN08}. At the highest power of
$6\, \Gamma_{21}$, population loss dominates and we see a
wide dip at line center with 30,000 times less absorption.

We have also done simulations at other control-laser
strengths. Apart from increased power broadening, the
general features of the strong-probe modifications are the
same.

\section{Results in a four-level system}
In a four-level system, there is enhanced absorption in the
middle of the EIT dip at line center. This narrow feature,
called Electromagnetically Induced Transparency and
Absorption (EITA) \cite{HCN05}, arises due to coherences
created between levels $\ket{1}$ and $\ket{4}$. The
coherences are generated by the simultaneous driving of
transitions $\ket{1} \leftrightarrow \ket{2}$, $\ket{2}
\leftrightarrow \ket{3}$, and $\ket{3} \leftrightarrow
\ket{4}$ by three different lasers. As in the three-level
case, all the population can be assumed to be in the ground
level in the weak-probe limit, i.e., $\rho_{11} = 1$, and
$\rho_{22} = \rho_{33} = \rho_{44} = 0$. This assumption
breaks down when the probe is strong. Series expansion of
$\rho_{12}$ in powers of $\Omega_{12}$ yields
\begin{equation}
\rho_{12}=\frac{-\frac{i}{2}\frac{\Omega_{12}}{\gamma_{12}}
\left[ 1-\frac{\Omega^2_{12}}{4\Gamma_{12}}\left( \frac{1}{\gamma_{12}}+\frac{1}{\gamma^*_{12}}\right)  +\ldots \right] }
{1+\frac{\Omega^2_{23}}{4\gamma_{12}\gamma_{13}} /
\left( 1+\frac{\Omega^2_{34}}{4\gamma_{13}\gamma_{14}}\right) } \, ,
 \label{EITAhigh}
\end{equation}
where
\begin{eqnarray}
\gamma_{12}&=&\Gamma_{21}/2+i\Delta_{12} \, , \nonumber \\
\gamma_{13}&=&(\Gamma_{32} + \Gamma_{34} +\Gamma_{35})/2 +i(\Delta_{12}+\Delta_{23}) \, , \nonumber \\
\gamma_{14}&=&\Gamma_{41}/2+i(\Delta_{12}+\Delta_{23}-\Delta_{34}) \, . \nonumber
\end{eqnarray}

\begin{figure}
\centering{\resizebox{0.95\columnwidth}{!}{\includegraphics{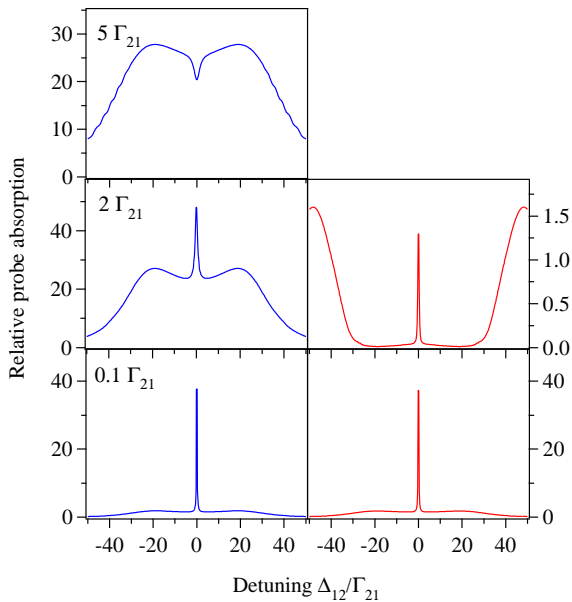}}}
\caption{Relative probe absorption [defined as
Im($-\rho_{12}$)] plotted as a function of probe detuning
for different values of probe Rabi frequency $\Omega_{12}$.
Both parameters are measured in units of $\Gamma_{21}$. The
two control lasers have strengths of $\Omega_{23}=2\pi
\times 10$~MHz and $\Omega_{34}=2\pi \times 5$~MHz. The
left-hand side is for a closed system and the right-hand
side is for an open system. The absorption at
$\Omega_{12}=5\, \Gamma_{12}$ for an open system is
negligible, and not shown.}
 \label{fourlevel}
\end{figure}

The numerical solutions of Im($-\rho_{12}$) from the
density-matrix equations for both closed systems (with
$\Gamma_{35} = 0$) and open systems are shown in Fig.\
\ref{fourlevel}. The calculations are done with the
control-laser strengths of $\Omega_{23} = 2\pi \times
10$~MHz ($1.7 \, \Gamma_{32}$) and $\Omega_{34} = 5$~MHz
($3.2 \, \Gamma_{34}$). As in the case of the three-level
system, the decay term governing $\rho_{13}$ is made same
by taking the decay rate from $\ket{3} \rightarrow \ket{4}$
for the closed system to be $\Gamma_{34}+\Gamma_{35}$. The
solutions are shown at $t=500$~$\mu$s.

In closed systems (shown on the left-hand side of the
figure), the phenomenon of enhanced absorption seen in the
weak-probe limit of $\Omega_{12}=0.1\, \Gamma_{21}$ becomes
less prominent when the Rabi frequency is increased to $2\,
\Gamma_{21}$. The linewidth of the feature also increases.
The feature inverts to an enhanced {\it transmission} dip
when the Rabi frequency becomes $5\, \Gamma_{21}$. In open
systems (shown on the right-hand side of the figure), the
line shape is similar to that of the closed system in the
weak-probe limit, as expected. However, when the probe Rabi
frequency is increased to $2\, \Gamma_{21}$, optical
pumping into the metastable level causes population loss
and reduces the absorption by a factor of 30. But the
overall EITA feature survives with the same linewidth. At
very high powers, population loss makes the absorption
disappear completely and is not shown.

\section{Conclusions}
In conclusion, we have seen that it is important to
consider the effects of a strong probe laser in several
applications of the phenomenon of EIT. The strong probe
laser causes population transfer and creates new pathways
for probe absorption. We solve the set of density-matrix
equations numerically to understand these effects in both
closed and open systems. For a closed three-level ladder
system, the EIT dip at line center disappears as the probe
laser becomes strong. In an open system, the absorption
level goes down because of population loss but the general
EIT feature at line center survives. For a four-level
system, there is a narrow enhanced-absorption peak at line
center. This converts to an EIT dip under strong-probe
conditions in a closed system. However, in an open system,
the enhanced-absorption feature survives even at high probe
powers albeit at a reduced absorption level. These
differences should be important in proposed clock
applications in four-level systems.

\begin{acknowledgments}
This work was supported by the Department of Science and
Technology, India. V.N. acknowledges support from the Homi
Bhabha Fellowship Council and K.P. from the Council of
Scientific and Industrial Research, India.
\end{acknowledgments}

%\bibliography{D:/papers/eitsubnat/eitrefs}

\begin{thebibliography}{15}
\expandafter\ifx\csname
natexlab\endcsname\relax\def\natexlab#1{#1}\fi
\expandafter\ifx\csname bibnamefont\endcsname\relax
  \def\bibnamefont#1{#1}\fi
\expandafter\ifx\csname bibfnamefont\endcsname\relax
  \def\bibfnamefont#1{#1}\fi
\expandafter\ifx\csname citenamefont\endcsname\relax
  \def\citenamefont#1{#1}\fi
\expandafter\ifx\csname url\endcsname\relax
  \def\url#1{\texttt{#1}}\fi
\expandafter\ifx\csname
urlprefix\endcsname\relax\def\urlprefix{URL }\fi
\providecommand{\bibinfo}[2]{#2}
\providecommand{\eprint}[2][]{\url{#2}}

\bibitem[{\citenamefont{Imamo\u{g}lu and
    Harris}(1989)}]{IMH89}
    \bibinfo{author}{\bibfnamefont{A.}~\bibnamefont{Imamo\u{g}lu}}
  \bibnamefont{and} \bibinfo{author}{\bibfnamefont{S.~E.}
  \bibnamefont{Harris}}, \bibinfo{journal}{Opt. Lett.}
  \textbf{\bibinfo{volume}{14}}, \bibinfo{pages}{1344} (\bibinfo{year}{1989}),
  \urlprefix\url{http://ol.osa.org/abstract.cfm?URI=ol-14-24-1344}.

\bibitem[{\citenamefont{Agarwal}(1991)}]{AGA91}
    \bibinfo{author}{\bibfnamefont{G.~S.}
    \bibnamefont{Agarwal}},
  \bibinfo{journal}{Phys. Rev. Lett.} \textbf{\bibinfo{volume}{67}},
  \bibinfo{pages}{980} (\bibinfo{year}{1991}).

\bibitem[{\citenamefont{Rapol and Natarajan}(2002)}]{RAN02}
    \bibinfo{author}{\bibfnamefont{U.~D.}
    \bibnamefont{Rapol}} \bibnamefont{and}
  \bibinfo{author}{\bibfnamefont{V.}~\bibnamefont{Natarajan}},
  \bibinfo{journal}{Europhys. Lett.} \textbf{\bibinfo{volume}{60}},
  \bibinfo{pages}{195–} (\bibinfo{year}{2002}).

\bibitem[{\citenamefont{Krishna
    et~al.}(2005)\citenamefont{Krishna, Pandey,
  Wasan, and Natarajan}}]{KPW05}
\bibinfo{author}{\bibfnamefont{A.}~\bibnamefont{Krishna}},
  \bibinfo{author}{\bibfnamefont{K.}~\bibnamefont{Pandey}},
  \bibinfo{author}{\bibfnamefont{A.}~\bibnamefont{Wasan}}, \bibnamefont{and}
  \bibinfo{author}{\bibfnamefont{V.}~\bibnamefont{Natarajan}},
  \bibinfo{journal}{Europhys. Lett.} \textbf{\bibinfo{volume}{72}},
  \bibinfo{pages}{221–} (\bibinfo{year}{2005}).

\bibitem[{\citenamefont{Harris
    et~al.}(1990)\citenamefont{Harris, Field, and
  Imamo\ifmmode~\breve{g}\else \u{g}\fi{}lu}}]{HFI90}
\bibinfo{author}{\bibfnamefont{S.~E.}
\bibnamefont{Harris}},
  \bibinfo{author}{\bibfnamefont{J.~E.} \bibnamefont{Field}}, \bibnamefont{and}
  \bibinfo{author}{\bibfnamefont{A.}~\bibnamefont{Imamo\ifmmode~\breve{g}\else
  \u{g}\fi{}lu}}, \bibinfo{journal}{Phys. Rev. Lett.}
  \textbf{\bibinfo{volume}{64}}, \bibinfo{pages}{1107} (\bibinfo{year}{1990}).

\bibitem[{\citenamefont{Zhang
    et~al.}(2009)\citenamefont{Zhang, Khadka,
  Anderson, and Xiao}}]{ZKA09}
\bibinfo{author}{\bibfnamefont{Y.}~\bibnamefont{Zhang}},
  \bibinfo{author}{\bibfnamefont{U.}~\bibnamefont{Khadka}},
  \bibinfo{author}{\bibfnamefont{B.}~\bibnamefont{Anderson}}, \bibnamefont{and}
  \bibinfo{author}{\bibfnamefont{M.}~\bibnamefont{Xiao}},
  \bibinfo{journal}{Physical Review Letters} \textbf{\bibinfo{volume}{102}},
  \bibinfo{eid}{013601} (pages~\bibinfo{numpages}{4}) (\bibinfo{year}{2009}),
  \urlprefix\url{http://link.aps.org/abstract/PRL/v102/e013601}.

\bibitem[{\citenamefont{Wielandy and
    Gaeta}(1998{\natexlab{a}})}]{WIG98}
    \bibinfo{author}{\bibfnamefont{S.}~\bibnamefont{Wielandy}}
    \bibnamefont{and}
  \bibinfo{author}{\bibfnamefont{A.~L.} \bibnamefont{Gaeta}},
  \bibinfo{journal}{Phys. Rev. Lett.} \textbf{\bibinfo{volume}{81}},
  \bibinfo{pages}{3359} (\bibinfo{year}{1998}{\natexlab{a}}).

\bibitem[{\citenamefont{Pandey
    et~al.}(2008)\citenamefont{Pandey, Wasan, and
  Natarajan}}]{PWN08}
\bibinfo{author}{\bibfnamefont{K.}~\bibnamefont{Pandey}},
  \bibinfo{author}{\bibfnamefont{A.}~\bibnamefont{Wasan}}, \bibnamefont{and}
  \bibinfo{author}{\bibfnamefont{V.}~\bibnamefont{Natarajan}},
  \bibinfo{journal}{J. Phys. B} \textbf{\bibinfo{volume}{41}},
  \bibinfo{pages}{225503 (8pp)} (\bibinfo{year}{2008}),
  \urlprefix\url{http://stacks.iop.org/0953-4075/41/225503}.

\bibitem[{\citenamefont{Hau et~al.}(1999)\citenamefont{Hau,
    Harris, Dutton, and
  Behroozi}}]{HHD99}
\bibinfo{author}{\bibfnamefont{L.~V.} \bibnamefont{Hau}},
  \bibinfo{author}{\bibfnamefont{S.~E.} \bibnamefont{Harris}},
  \bibinfo{author}{\bibfnamefont{Z.}~\bibnamefont{Dutton}}, \bibnamefont{and}
  \bibinfo{author}{\bibfnamefont{C.~H.} \bibnamefont{Behroozi}},
  \bibinfo{journal}{Nature (London)} \textbf{\bibinfo{volume}{397}},
  \bibinfo{pages}{594} (\bibinfo{year}{1999}).

\bibitem[{\citenamefont{Hong
    et~al.}(2005)\citenamefont{Hong, Cramer,
  Nagourney, and Fortson}}]{HCN05}
\bibinfo{author}{\bibfnamefont{T.}~\bibnamefont{Hong}},
  \bibinfo{author}{\bibfnamefont{C.}~\bibnamefont{Cramer}},
  \bibinfo{author}{\bibfnamefont{W.}~\bibnamefont{Nagourney}},
  \bibnamefont{and} \bibinfo{author}{\bibfnamefont{E.~N.}
  \bibnamefont{Fortson}}, \bibinfo{journal}{Phys. Rev. Lett.}
  \textbf{\bibinfo{volume}{94}}, \bibinfo{pages}{050801}
  (\bibinfo{year}{2005}).

\bibitem[{\citenamefont{Santra
    et~al.}(2005)\citenamefont{Santra, Arimondo,
  Ido, Greene, and Ye}}]{SAI05}
\bibinfo{author}{\bibfnamefont{R.}~\bibnamefont{Santra}},
  \bibinfo{author}{\bibfnamefont{E.}~\bibnamefont{Arimondo}},
  \bibinfo{author}{\bibfnamefont{T.}~\bibnamefont{Ido}},
  \bibinfo{author}{\bibfnamefont{C.~H.} \bibnamefont{Greene}},
  \bibnamefont{and} \bibinfo{author}{\bibfnamefont{J.}~\bibnamefont{Ye}},
  \bibinfo{journal}{Phys. Rev. Lett.} \textbf{\bibinfo{volume}{94}},
  \bibinfo{pages}{173002} (\bibinfo{year}{2005}).

\bibitem[{\citenamefont{Taichenachev
    et~al.}(2006)\citenamefont{Taichenachev,
  Yudin, Oates, Hoyt, Barber, and Hollberg}}]{TYO06}
\bibinfo{author}{\bibfnamefont{A.~V.}
\bibnamefont{Taichenachev}},
  \bibinfo{author}{\bibfnamefont{V.~I.} \bibnamefont{Yudin}},
  \bibinfo{author}{\bibfnamefont{C.~W.} \bibnamefont{Oates}},
  \bibinfo{author}{\bibfnamefont{C.~W.} \bibnamefont{Hoyt}},
  \bibinfo{author}{\bibfnamefont{Z.~W.} \bibnamefont{Barber}},
  \bibnamefont{and} \bibinfo{author}{\bibfnamefont{L.}~\bibnamefont{Hollberg}},
  \bibinfo{journal}{Phys. Rev. Lett.} \textbf{\bibinfo{volume}{96}},
  \bibinfo{pages}{083001} (\bibinfo{year}{2006}).

\bibitem[{\citenamefont{Wielandy and
    Gaeta}(1998{\natexlab{b}})}]{WIG98a}
    \bibinfo{author}{\bibfnamefont{S.}~\bibnamefont{Wielandy}}
    \bibnamefont{and}
  \bibinfo{author}{\bibfnamefont{A.~L.} \bibnamefont{Gaeta}},
  \bibinfo{journal}{Phys. Rev. A} \textbf{\bibinfo{volume}{58}},
  \bibinfo{pages}{2500} (\bibinfo{year}{1998}{\natexlab{b}}).

\bibitem[{\citenamefont{Pandey and
    Natarajan}(2008)}]{PAN08}
    \bibinfo{author}{\bibfnamefont{K.}~\bibnamefont{Pandey}}
    \bibnamefont{and}
  \bibinfo{author}{\bibfnamefont{V.}~\bibnamefont{Natarajan}},
  \bibinfo{journal}{J. Phys. B} \textbf{\bibinfo{volume}{41}},
  \bibinfo{pages}{185504 (4pp)} (\bibinfo{year}{2008}),
  \urlprefix\url{http://stacks.iop.org/0953-4075/41/185504}.

\bibitem[{\citenamefont{Loftus
    et~al.}(2002)\citenamefont{Loftus, Bochinski,
  and Mossberg}}]{LBM02}
\bibinfo{author}{\bibfnamefont{T.}~\bibnamefont{Loftus}},
  \bibinfo{author}{\bibfnamefont{J.~R.} \bibnamefont{Bochinski}},
  \bibnamefont{and} \bibinfo{author}{\bibfnamefont{T.~W.}
  \bibnamefont{Mossberg}}, \bibinfo{journal}{Phys. Rev. A}
  \textbf{\bibinfo{volume}{66}}, \bibinfo{pages}{013411}
  (\bibinfo{year}{2002}).

\end{thebibliography}

\end{document}